\documentclass[12pt]{article}
\begin{document}
\begin{center}
\textbf{ CAN  A  KASNER  UNIVERSE  WITH  A  VISCOUS  COSMOLOGICAL  FLUID  BE  ANISOTROPIC?}\\

\bigskip
\bigskip

\bigskip

I. Brevik\footnote{E-mail address: iver.h.brevik@mtf.ntnu.no}

\bigskip

Division of Applied Mechanics, Norwegian University of Science and Technology,\\
N-7491 Trondheim, Norway\\

\bigskip
\bigskip

S. V. Pettersen\footnote{E-mail address: svp@fysel.ntnu.no}\\

\bigskip

Department of Physical Electronics, Norwegian University of Science and Technology,\\
N-7491 Trondheim, Norway\\

\bigskip
\bigskip

\bigskip
\bigskip

March 2000
\end{center}

\bigskip

\begin{abstract}
A Bianchi type -I metric of Kasner form is considered, when the space is filled with a viscous fluid. Whereas an ideal (nonviscous) fluid permits the Kasner metric to be anisotropic provided that the fluid satisfies the Zel'dovich equation of state, the viscous fluid does not permit the Kasner metric to be anisotropic at all. In the latter case, we calculate the Kasner (isotropic) metric expressed by the fluid's density, pressure, and bulk viscosity, at some chosen  instant $t=t_0$. The equation of state is also calculated. The present paper is related to a recent Comment of Cataldo and del Campo [Phys. Rev. D, scheduled to April 15, 2000], on a previous work of the present authors [Phys. Rev. D {\bf 56}, 3322 (1997)].

\end{abstract}
\newpage
Consider a cosmic fluid, endowed with a bulk viscosity $\zeta$ and a shear viscosity $\eta$. In a homogeneous (possibly anisotropic) space, $\zeta$ and $\eta$ are independent of position, but will in general depend on time. If $U^\mu =(U^0, U^i)$ is the fluid's four-velocity, the energy-momentum tensor is
\begin{equation}
T_{\mu\nu}=\rho U_\mu U_\nu+(p-\zeta\theta)h_{\mu\nu}-2\eta \sigma_{\mu\nu}.
\end{equation}
\label{1}
Here $h_{\mu\nu}=g_{\mu\nu}+U_\mu U_\nu$ is the projection tensor, $\theta=\theta_\mu^\mu ={U^\mu}_{;\mu}$ is the scalar expansion, $\theta_{\mu\nu}=\frac{1}{2}(U_{\mu;\alpha}h_\nu^\alpha+U_{\nu;\alpha}h_\mu^\alpha)$ is the expansion tensor, and $\sigma_{\mu\nu}=\theta_{\mu\nu}-\frac{1}{3}h_{\mu\nu} \theta$ is the shear tensor.

Assume now that the metric of the background space, in which the cosmic fluid resides, is of the Kasner form:
\begin{equation}
ds^2=-dt^2+t^{2p_1}\,dx^2+t^{2p_2}\,dy^2+t^{2p_3}\,dz^2.
\end{equation}
\label{2}
The three numbers $p_1, p_2, p_3$ are required to be constants. The Kasner metric is a subclass of the Bianchi type-I  metrics. The space is anisotropic if at least two of the three $p_i$ are different. Defining numbers $S$ and $Q$ by $S=\sum_{i=1}^3 p_i$ and  $Q=\sum_{i=1}^3 p_i^2$ we have, for a strict Kasner universe in the classical sense corresponding to a pure {\it vacuum}, that $S=Q=1$. Once a real, in general viscous fluid is present, however, these simple relationships are lost.

Consider next the Einstein equations. With the cosmological constant $\Lambda$ set equal to zero, and with the notation $\kappa=8\pi G$, we obtain from $R_{\mu\nu}=\kappa(T_{\mu\nu}-\frac{1}{2}g_{\mu\nu} T_\alpha ^\alpha)$ the two equations
\begin{equation}
S-Q+\frac{3}{2}\kappa t \zeta S=\frac{1}{2}\kappa t^2(\rho+3p),
\end{equation}
\label{3}
\begin{equation}
p_i(1-S-2\kappa t \eta)+\frac{1}{2}\kappa t (\zeta+\frac{4}{3}\eta)S
=-\frac{1}{2}\kappa t^2(\rho-p).
\end{equation}
\label{4}
These are the governing field equations. Adding the three equations (4) we get
\begin{equation}
2S(S-1)=3\kappa t\zeta S+3\kappa t^2(\rho-p),
\end{equation}
\label{5}
whereas from Eqs. (3) and (5) we obtain the simple equation
\begin{equation}
S^2-Q=2\kappa t^2 \rho.
\end{equation}
\label{6}
Faced with these equations, we see that there are several factors determining the state of the cosmic fluid:

(i)  The viscosity coefficients $\zeta$ and $\eta$.

(ii)  The constants $S$ and $Q$, which may be regarded as input parameters in the governing equations.

(iii)  In turn, $S$ and $Q$ are for a viscous fluid determined by the energy density $\rho=\rho_0$ and pressure $p=p_0$ evaluated at some chosen instant $t=t_0$ which, in terms of an appropriately scaled time variable, will be taken at $t_0=1$. The time variations of $\rho=\rho(t)$ and $p=p(t)$ follow from the governing equations themselves. By the same token, the time variations of $\zeta=\zeta(t)$ and $\eta=\eta(t)$ follow. Once the proportionaly constants $ \{\rho_0, p_0, \zeta_0, \eta_0\}$ are known, the Kasner parameters $p_i$ and, as mentioned, also $S$ and $Q$, are known.

(iv)  The possible equation of state for the fluid. Conventionally, the equation of state is written in the form $p=(\gamma-1)\rho$, where $\gamma$ is a constant. An important point, which can be resolved only after a detailed analysis of the formalism, is to what extent the equation of state is fixed once $\{\rho_0, p_0, \zeta_0, \eta_0 \}$ are known.

In a previous paper \cite{brevik97} we analysed, on basis of the governing equations above, the possible equation of state for the fluid. A recent Comment of Cataldo and del Campo \cite{cataldo00} claims that one of the options discussed in \cite{brevik97} (actually a singular option), involving a viscous fluid residing in an anisotropic space, is not physically realizable because it runs into conflict with the {\it dominant energy condition}. The remarks of Cataldo and del Campo are interesting, since they seem to narrow down, on physical grounds, the range of applicability of the anisotropic Kasner metric for a realistic cosmic fluid. In turn, this may even lead to consequences for the applicability of anisotropic viscous universe models in general. We have found it desirable, therefore, to discuss this point in some detail, from a more broad perspective than in Ref. \cite{cataldo00}. This is the purpose of the present paper.

Let us first consider the physical meaning of the dominant energy condition. This condition is usually expressed mathematically by saying that in a local rest orthonormal frame the magnitude of the stress components $T_{\hat{i}\hat{k}}$ are always less than or equal to the energy density component \cite{hawking73}:
\begin{equation}
|T_{\hat{i}\hat{k}}| \leq \rho.
\end{equation}
\label{7}
This ought to be contrasted with the {\it weak} energy condition, which says that the energy density in an orthonormal rest inertial frame is always non-negative, $\rho \geq 0$. The dominant energy condition is the weak energy condition with the additional requirement that the stress components - in practice the diagonal components of $T_{\hat{i} \hat{k}}$ - should not exceed the energy density. Moreover, since the velocity of sound in a fluid equals the square root of the adiabatic derivative of $p$ with respect to $\rho$, the dominant condition is strongly related to the physical property of a sound wave that it cannot propagate faster than light.

In the present case, $\sigma_{ii}=(p_i-\frac{1}{3}S)t^{2p_i-1}$ (no sum over $i$) and $\theta=S/t$ \cite{brevik97}. From Eqs. (1) and (7) we then derive, in the local rest orthonormal frame,
\begin{equation}
\left| p-\frac{\zeta S}{t}-\frac{2\eta}{t} (p_i-\frac{1}{3}S) \right| \leq \rho.
\end{equation}
\label{8}
This is the viscous generalization of the dominant energy condition. For an ideal fluid, Eq. (8) reduces to the well known condition $ |p| \leq \rho$.

Let us now consider the time dependence of the physical quantities. From Eqs. (3)-(6) it is evident that $\rho(t)$ and $p(t)$ are proportional to $t^{-2}$, whereas $\zeta(t)$ and $\eta(t)$ are proportional to $t^{-1}$. We write
\begin{equation}
\rho(t)=\rho_0 t^{-2},~~~~p(t)=p_0 t^{-2},
\end{equation}
\label{9}
\begin{equation}
\zeta(t)=\zeta_0 t^{-1},~~~~\eta(t)=\zeta_0 t^{-1},
\end{equation}
\label{10}
where $\{\rho_0, p_0, \zeta_0, \eta_0\}$ are the proportionality constants. The proportionalities when written in this form imply that the instant $t=0$ is taken to be a singular point. This is in accordance with the fact that the Kasner metric when written such as in Eq. (2) implies the true singularity (i. e., the divergence of the contracted Riemann tensor components) to occur at $t=0$ (cf., for instance, Ref. \cite{ryan75}). We can now write the field equations (3)-(6) as
\begin{equation}
S-Q+\frac{3}{2}\kappa \zeta_0 S=\frac{1}{2}\kappa (\rho_0+3p_0),
\end{equation}
\label{11}
\begin{equation}
p_i(1-S-2\kappa  \eta_0)+\frac{1}{2}\kappa (\zeta_0+\frac{4}{3}\eta_0)S=-\frac{1}{2}\kappa
(\rho_0-p_0),
\end{equation}
\label{12}
\begin{equation}
2S(S-1)=3\kappa  \zeta_0S+3\kappa (\rho_0-p_0),
\end{equation}
\label{13}
\begin{equation}
S^2-Q=2\kappa \rho_0.
\end{equation}
\label{14}
These equations contain time-independent terms only. The dominant energy condition (8) becomes correspondingly
\begin{equation}
\left| p_0-\zeta_0 S -2\eta_0(p_i-\frac{1}{3}S) \right| \leq \rho_0.
\end{equation}
\label{15}
After having established the governing equations and the dominant energy condition, we proceed to categorize the different options for the equation of state for the fluid. We follow essentially the same line of approach as in Ref. \cite{brevik97}, and consider the ideal fluid first.

{\it Ideal fluid}.   When $\zeta=\eta=0$ we consider, as first option, the case
\begin{equation}
p_0=\rho_0,
\end{equation}
\label{16}
i. e., a Zel'dovich fluid. This is a maximally stiff equation of state; the velocity of sound is equal to the velocity of light. From Eqs. (11)-(14) we  obtain
\begin{equation}
S=1,~~~~Q=1-2\kappa \rho_0.
\end{equation}
\label{17}
Thus two of the Kasner parameters, say $p_1$ and $p_2$, can be found as solutions of the second-degree equation
\begin{equation}
p_1^2+p_1p_2+p_2^2-p_1-p_2+\kappa  \rho_0=0,
\end{equation}
\label{18}
which describes a conic section in the $p_1 p_2$ plane \cite{korn68}. If
\begin{equation}
\kappa \rho_0 < \frac{1}{3},
\end{equation}
\label{19}
the curve is an ellipse. In this case there is a continuous family of values for $p_1$ and $p_2$, permitting the Kasner space to be anisotropic. The third Kasner parameter follows as $p_3=1-p_1-p_2$. The condition (19) is satisfied under usual physical circumstances. If this condition is {\it not} satisfied, there is no real locus (imaginary ellipse).

It is remarkable that the initial Kasner form of the metric is, for an ideal fluid, compatible with the Zel'dovich state equation only. For any state equation different from $p_0=\rho_0$ (and this is our second option), we see from  Eq. (4)  that all the three $p_i$'s have to be equal. We then find, putting $p_1=p_2=p_3 \equiv a$,
\begin{equation}
a=\frac{1}{6}\left[ 1+\sqrt{1+6\kappa (\rho_0-p_0)}\right].
\end{equation}
\label{20}
That is, knowledge about the proportionality constants $\rho_0$ and $p_0$  suffices to calculate the three equal Kasner parameters in the isotropic space. (The reason why the case $p_0=\rho_0$ in Eq. (20) yields $a=\frac{1}{3}$, instead of the family of solutions given in Eq. (18), is that we are dealing with two different options.)

{\it Viscous fluid.}  This is for us of main interest. Consider first the exceptional case when
\begin{equation}
S=1-2\kappa  \eta_0.
\end{equation}
\label{21}
From Eq. (12) it might seem possible, at first sight, that in this case anisotropic space is permitted. However, {\it if} $\rho_0 \geq p_0$, there is a conflict between the remaining terms in the equation since the left hand side (involving positive viscosity coefficients) is positive whereas the right hand side becomes negative.

Now, the careful reader may ask: why do we know that $ \rho_0 \geq p_0 $? In particular, does this condition follow from the dominant energy condition alone, as stated in \cite{cataldo00}?  Let us return to Eq. (15): we see herefrom that $ \rho_0-p_0 \geq -\zeta_0 S-2\eta_0(p_i-\frac{1}{3}S)$. If $p_i \geq \frac{1}{3}S$ this is a {\it weaker} condition than the condition $\rho_0-p_0 \geq 0$. Inserting this inequality into Eq. (12) we find that $S \leq p_i$, which is equivalent to  $ p_2+p_3 \leq 0$, $p_1+p_3 \leq 0$, $p_1+p_2 \leq 0$. These inequalities are actually {\it acceptable}. The dominant energy condition alone is thus not sufficient. What really makes us accept the restriction $\rho_0-p_0 \geq 0 $ is rather the physical property mentioned above: a sound wave cannot propagate faster than light.  We  end up with the conclusion that the option (21) has to be abandoned, thus in agreement with \cite{cataldo00}, although our reasoning is different from that of \cite{cataldo00}. 

In the {\it isotropic} Kasner space we now obtain, from Eq. (12),
\begin{equation}
a=\frac{1}{6}\left[ 1+\frac{3}{2}\kappa  \zeta_0+\sqrt{(1+\frac{3}{2}\kappa  \zeta_0)^2+6\kappa 
 (\rho_0-p_0)}~~\right],
\end{equation}
\label{22}
showing that the isotropic Kasner metric (2) is completely determined when  $\rho_0$ and $p_0$, and the bulk viscosity $\zeta_0$, are known. It is notable that the {\it shear} viscosity $\eta_0$ is absent in Eq. (22); this is in accordance with the fact that the shear viscosity is a concept related to an anisotropic state of motion for a fluid. If the fluid is ideal, Eq. (22) is seen to reduce to Eq. (20).

As for the equation of state for the fluid, we note from  the governing equations that
\begin{equation}
\kappa \rho_0=3a^2,
\end{equation}
\label{23}
\begin{equation}
\kappa p_0=2a-3a^2+3a\kappa  \zeta_0.
\end{equation}
\label{24}
That means that we are allowed to write the equation of state in the conventional form $p=(\gamma-1)\rho$, or $p_0=(\gamma -1)\rho_0$, where the constant $\gamma$ is
\begin{equation}
\gamma=\frac{2+3\kappa  \zeta_0}{3a}.
\end{equation}
\label{25}
It ought finally to be emphasized that we have been considering the Kasner subclass of Bianchi type-I metrics only. We cannot, on basis of the remarks above, make any conclusion about the existence of viscous cosmic fluids in the whole Bianchi type-I class.

\bigskip

{\bf Acknowledgment}

\bigskip

It is a pleasure to acknowledge valuable remarks by Professor {\O}yvind Gr{\o}n on the present manuscript.

\end{document}